\documentclass[12pt]{article}
\usepackage{times}
\usepackage{changepage} 
\usepackage{hyperref}
\usepackage[T1]{fontenc} 
\usepackage{breakcites} 
\usepackage{url} 

\newcommand{\itemb}{\begin{itemize}}
\newcommand{\iteme}{\end{itemize}}

\newcommand{\enumb}{\begin{enumerate}}
\newcommand{\enume}{\end{enumerate}}

\newcommand{\smallest}[1]{\vspace{6pt}\noindent {\bf  {#1}.}}

\newcommand{\com}[1]{}
\usepackage{xcolor}

\newcommand{\ocd}{obsessive-compulsive disorder}

\newcommand{\indep}{independent}

\newcommand{\antip}{antipsychotic}

\newcommand{\xtblty}{excitability}

\newcommand{\desensit}{desensitization}

\newcommand{\hthal}{hypothalamus}

\newcommand{\amg}{amygdala}

\newcommand{\sympa}{{{sympathetic}}}

\newcommand{\infl}{{{inflammation}}}

\newcommand{\catwo}{{{Ca}\textsuperscript{{\raisebox{-1pt}{$\scriptstyle{2+}$}}}}}

\newcommand{\kplus}{{{K}\textsuperscript{+}}}

\newcommand{\naplus}{{{Na}\textsuperscript{+}}}

\newcommand{\thn}{T*OCD} 

\usepackage{fancyhdr}
\title{A CRH-HCN Theory of Obsessive-Compulsive Disorder (OCD)}
\author{
{Ari Rappoport}\\ 
The Hebrew University of Jerusalem, Israel\\
ari.rappoport@mail.huji.ac.il
}
\date{January 2024}

\begin{document}
\maketitle
\section*{Abstract}
I present the first complete theory of OCD. OCD occurs when excessive CRH is released in the prefrontal cortex, activating cAMP. cAMP is a major inducer of HCN channels, which promote repeated neural firing. The combination of CRH, which is strongly accociated with stress, and repeated firing that cannot be controlled, explains all of the features of OCD, including obsessions and compulsions of all kinds. 
\section*{Theory}

\itemb

\item I present a {\bf CRH-HCN theory of \ocd\ (OCD) (\thn)}. \thn\ is the first complete theory of OCD, explaining its etiology, symptoms, pathophysiology, and treatment. 

\item {\bf The core cause of OCD is chronically increased release of corticotropin-releasing hormone (CRH), coupled with chronic activation of hyperpolarization-activated cyclic nucleotide-gated (HCN) channels}. HCN channels explain the mechanisms behind OCD symptoms, while CRH excites HCN channels and also explains the association with stress. 


\item CRH is the major brain agent released following deviations from homeostasis (i.e., stress), conveying the need for restoration responses. It acts on two receptors, CRH1 and CRH2, with higher affinity for CRH1. Both receptors are G protein-coupled receptors that mainly bind Gs to induce cAMP, but also Gi/o and Gq. 

\item Strong or chronic stress during sensitive developmental periods can induce persistent alterations to the stress response system. Such alterations commonly involve chronically increased release of stress-related agents, and can be heritable (e.g., via epigenetics).  

\item According to \thn, {\bf OCD occurs when chronically released CRH induces excessive activation of CRH1 and possibly CRH2}. This yields excessive cAMP, which induces chronic HCN channel activity as explained below. Alternatively, OCD can occur due to {\bf CRH-\indep\ chronic activation of HCN channels}. 

\item HCN channels are activated by small hyperpolarization and inactivated by small polarization. They are permeable to \naplus\ and \kplus, and when open yield an excitatory current (Ih) that increases polarization close to the action potential threshold. Thus, their opening facilitates {\bf intrinsic slow cellular oscillations}. HCN channels are crucial to heart rhythmic activity, and are strongly expressed in the brain, including in stress-related areas, cortical response neurons (layer 5 pyramidal neurons that project outside of cortex), the basal ganglia (BG), and the thalamus.

\item cAMP is a major stimulator of HCN channel opening. Thus, chronic CRH signaling induces chronic activation of HCN channels, which promotes repeated neural firing in stress-related areas (i.e., areas with CRHR expression). 

\item Activation of CRH1 by CRH normally leads to its endocytosis and \desensit. However, in the presence of chronic CRH, this does not prevent chronic cAMP synthesis, because (i) if CRH does not bind CRH1, it accumulates and binds the lower affinity CRH2, which also stimulates cAMP, (ii) endocytosed CRH1 and CRH2 continue producing cAMP via lysosomes, and (iii) chronic CRH activates CRH1 receptors again when they are retransported to the plasma membrane. 

\item According to \thn, {\bf the obsessions exhibited by OCD patients are due to chronic neural firing driven by cAMP and HCNs}. Due to the increased push for HCN activation, it is very difficult for the brain to suppress repeated activations in paths that express CRHR-HCN. Thus, stress-related thoughts and actions are continuously reactivated without control. 
\item OCD is strongly associated with a phenomenon called {\bf `not just right'}, a feeling of {\bf incompleteness} commonly manifesting as confirmation questions. This phenomenon results from the same chronic CRH-HCN process that drives obsessions. The need to answer a question/challenge (even simple questions), which is normally conveyed by CRH, continues reverberating due to HCN-induced oscillations, even after a response has been reached. {\bf This also explains continuous rechecking behavior in OCD}. 

\item Many OCD patients exhibit {\bf strong binary views} and {\bf rigid personality}. These occur because response neurons are sometimes not re-activated, making the person very confident in these responses (since patients are used to uncertainty). A lack of re-activation often results from inputs from valence areas (e.g., the \amg), which provide strong excitation of cortical response neurons, especially to frontal layer 2 pyramidal cells, which are also the ones where CRH-HCN expression is highest. {\bf This explains why the rigid personality in OCD is often related to moral issues}.  

\item OCD {\bf compulsions} are a form of self-treatment in which patients voluntarily activate neurons. They do this because {\bf neural activation induces moderate \catwo\ influx, which temporarily inhibits cAMP and relieves obsessions and the feeling of uncertainty}. With repeated execution, compulsions get desensitized, gradually losing their ability to stimulate \catwo\ influx and prompting patients to increase compulsion complexity and/or number of repeats. 

\item OCD behaviors related to {\bf cleanliness} (e.g., hand washing) are very common. This occurs due to two reasons. First, CRH activates mast cells, and induces the release of histamine (both directly and via mast cells). Mast cells are innate immune cells activated by various pathogens. Histamine induces a feeling of itch, and a large variety of agents released by mast cells report the presence of pathogens and/or \infl\ to the brain. Second, CRH yields ACTH release in skin. This can induce the production of beta-endorphin, which is known to induce itch via mu-opioid receptors. {\bf The chronic CRH in OCD yields chronic activation of circuits conveying skin-related pathogen threats, and an uncomfortable feeling}. 
Hand-washing can  temporarily relieve the problem as in other compulsions, and via the activation and desensitization of TRPV1 channels. 

\item A similar account also explains {\bf trichotillomania} and {\bf excoriation}, OCD sub-types that involve hair pulling and skin picking, respectively, and nail picking in OCD. CRH1, mast cells and histamine are involved in hair growth.
\item {\bf Hoarding} is another OCD subtype. It is explained by the feeling of incompleteness applied to object use (`Am I finished with this object?' `I'm not sure. I may need it in the future, so I won't throw it away').  

\item {\bf From the patient perspective, OCD behavior is a rational response to the inputs that their brain receives}. The problem in OCD is that these inputs are not connected to the real state of the organism, but result from erroneous release of a molecule. However, patients experience such inputs exactly as they experience real ones. This explains the (small) subset of patients who do not show {\bf `insight'} as to their symptoms. 

\item Beyond the biological logic related to the known effects of CRH and HCN channels, there is {\bf empirical evidence} supporting \thn. CRH1 recruits the \sympa\ nervous system (SNS) and the hypothalamus-pituitary-adrenal (HPA) stress axis (resulting in the release of the human glucocorticoid, cortisol).
OCD patients show increased CRH, ACTH, and cortisol (indicating chronic CRH), reduced sensitivity to CRH challenge (showing chronic release), reduced cortisol-induced negative feedback (which is partly done via CRH1), reduced prepulse inhibition of the startle reflex (indicating primed CRH), and chronically higher activity in brain areas expressing HCN (`CSTC loops'). In addition, OCD is associated with genetic mutations in HCN and PDE (cAMP suppressor) genes, with increased allergies (indicating mast cell, histamine), and with trauma (supporting stress-related etiology). There is no contradictory evidence.

\item OCD patients have {\bf trouble falling asleep}. This provides additional evidence for \thn, since CRH1 stimulates the SNS and cortical norepinephrine (NEP), both of which oppose sleep. 

\item \thn\ is also supported by known OCD comorbidities. There is high comorbidity with {\bf anxiety}, which is known to be linked to CRH. 

\item OCD shows high comorbidity with {\bf panic disorder}. \thn\ explains this by noting that the enhanced HCN channel and SNS activity in OCD are expected to increase the likelihood of panic attacks, due to their effects on the heart. 

\item Supporting SNS dysregulation, OCD patients show higher rates of {\bf cardiovascular disease}. 

\item {\bf Traumatic brain injury (TBI)} can induce a variety of psychiatric conditions, including OCD. TBI increases CRH and affects HCN channels. 

\iteme

\section*{Treatment}
\itemb

\item {\bf Cognitive behavioral therapy (CBT)} is helpful in OCD, especially when combined with medication treatment. CBT teaches patients to reduce the attention given to OCD paths, thereby reducing their \xtblty\ and enhancing non-OCD paths.  

\item The best drug treatment for OCD would be to {\bf reduce CRH release}. Unfortunately, at present there is no selective drug that directly has this effect. 

\item {\bf Selective serotonin reuptake inhibitors (SSRIs)} increase brain serotonin (SER) levels. In high doses, SER opposes the low CRH mode and CRH1 signaling. Moreover, SER reduces HCN currents via PKC. Thus, SSRIs can significantly help OCD. However, in lower doses, SER cooperates with the high CRH mode and CRH2 signaling. Thus, SSRIs only help when provided at the right doses (especially when coupled with antagonists of the CRH2-related SER2 receptors, such as atypical \antip s). In practice, this only works in a subset of patients, because it is very difficult to continuously calibrate the doses required, and because the effect of SSRIs is non-selective and indirect. 

\item Alpha2 adrenergic (NEPa2) receptor agonists cross the blood-brain barrier and oppose brain HCN channel activation, and hence might be beneficial in OCD. {\bf NEPa2 agonists such as guanfacine and clonidine} are routinely used to treat ADHD, a disorder with substantial comorbidity with OCD. There are some indications from a small number of case reports that these drugs are helpful in OCD. {\bf Usage of these drugs to treat OCD, alone or with SSRIs, should be seriously explored}.

\item {\bf Guanfacine is preferable over clonidine}, because it is more selective to the NEPa2a subtype, which  has a higher expression in stress-related areas such as the prefrontal cortex, \amg, \hthal, and locus coeruleus. 

\item When used for ADHD, NEPa2 agonists commonly induce sleepiness and sedation due to suppression of NEP (and possibly epinephrine (EPI)) release. {\bf In OCD, reducing NEP and EPI may actually be desirable}, because the chronic CRH problem in OCD induces chronic NEP and EPI release (see above). 

\item SSRIs and guanfacine can be used alone or in combination. {\bf For patients with minimal anxiety symptoms (i.e., with relatively `pure' OCD symptoms), guanfacine may suffice}, due to its direct suppression of HCN channels. SSRIs may be essential for addressing patients with higher anxiety symptoms. 
\iteme

\section*{References}

Available upon request. 
\end{document}